\documentclass[aps,prl,twocolumn,superscriptaddress]{revtex4}
\usepackage[dvipdfmx]{graphicx}% To use PDF
\usepackage{dcolumn}% Align table columns on decimal point
\usepackage{bm}% bold math
\usepackage{amsmath}
\usepackage{amssymb}
\usepackage{latexsym}
\usepackage{epsfig}
\usepackage{amsbsy}
\usepackage{array}
\usepackage{amssymb}
\usepackage{setspace}
\usepackage{bm}
\usepackage{epstopdf}
\usepackage{color}
\usepackage[colorlinks=true,urlcolor=blue,linkcolor=blue,citecolor=blue]{hyperref}% add hypertext capabilities

\def\sint{\ifmmode{- \!\!\!\!\!\! \int}
    \else{\hbox{$- \!\!\!\! \int \ $}}\fi}

\begin{document}

%\preprint{Physical Review Letters}

\title{Field-induced Topological Hall effect in antiferromagnetic axion insulator candidate EuIn$_2$As$_2$}% Force line breaks with \\

\author{J. Yan}
\email{jianyan@issp.u-tokyo.ac.jp}
\affiliation{The Institute for Solid State Physics, The University of Tokyo, Kashiwa, 277-8581, Japan}
\affiliation{Key Laboratory of Materials Physics, Institute of Solid
State Physics, HFIPS, Chinese Academy of Sciences, Hefei 230031, China}
\affiliation{University of Science and Technology of China, Hefei, 230026, China}
\author{Z. Z. Jiang}
\affiliation{Key Laboratory of Materials Physics, Institute of Solid
State Physics, HFIPS, Chinese Academy of Sciences, Hefei 230031, China}
\affiliation{University of Science and Technology of China, Hefei, 230026, China}
\author{R. C. Xiao}
\affiliation{Institutes of Physical Science and Information Technology, Anhui University,
Hefei, 230601, China}
\author{W. J. Lu}
\affiliation{Key Laboratory of Materials Physics, Institute of Solid
State Physics, HFIPS, Chinese Academy of Sciences, Hefei 230031, China}
\author{W. H. Song}
\affiliation{Key Laboratory of Materials Physics, Institute of Solid
State Physics, HFIPS, Chinese Academy of Sciences, Hefei 230031, China}
\author{X. B. Zhu}
\affiliation{Key Laboratory of Materials Physics, Institute of Solid
State Physics, HFIPS, Chinese Academy of Sciences, Hefei 230031, China}
\author{X. Luo}
\affiliation{Key Laboratory of Materials Physics, Institute of Solid
State Physics, HFIPS, Chinese Academy of Sciences, Hefei 230031, China}
\author{Y. P. Sun}
\email{ypsun@issp.ac.cn}
\affiliation{Anhui Province Key Laboratory of Condensed Matter Physics at Extreme Conditions, High Magnetic Field Laboratory, HFIPS, Chinese Academy of
Sciences, Hefei 230031, China}
\affiliation{Key Laboratory of Materials Physics, Institute of Solid
State Physics, HFIPS, Chinese Academy of Sciences, Hefei 230031, China}
\affiliation{Collaborative Innovation Center of Microstructures,
Nanjing University, Nanjing 210093, China}
\author{M. Yamashita}
\email{my@issp.u-tokyo.ac.jp}
\affiliation{The Institute for Solid State Physics, The University of Tokyo, Kashiwa, 277-8581, Japan}

\begin{abstract}
The magnetic topological materials have attracted significant attention due to their potential realization of variety of novel quantum phenomena. EuIn$_2$As$_2$ has recently been theoretically recognized as a long awaited intrinsic antiferromagnetic bulk axion insulator. However, the experimental study on transport properties arising from the topological states in this material is scarce. In this paper, we perform the detailed magnetoresistance (MR) and Hall measurements to study the magnetotransport properties of this material.
We find that the transport is strongly influenced by the spin configuration of the Eu moments from the concomitant change in the field dependence of the MR and that of the magnetization below the N{\'e}el temperature.
%The concomitant field dependence of the MR and that of the magnetization observed below the N{\'e}el temperature provide evidence that the transport phenomena is strongly influenced by the spin configuration of the Eu moments in this system.
Most importantly, an anomalous Hall effect (AHE) and a large topological Hall effect (THE) are observed. We suggest that the AHE is originated from a nonvanishing net Berry curvature due to the helical spin structure and that the THE is attributed to the formation of a noncoplanar spin texture with a finite scalar spin chirality induced by the external magnetic field in EuIn$_2$As$_2$. Our studies provide a platform to understand the influence of the interplay between the topology of electronic bands and the field-induced magnetic structure on magnetoelectric transport properties. In addition, our observations give a hint to realize axion insulator states and high-order topological insulator states through manipulating the magnetic state of EuIn$_2$As$_2$.
\end{abstract}

%\pacs{75.80.+q, 77.65.-j}


\maketitle

\section{Introduction}  %This should be cancelled for PRL and APL.

The success of the topological band theory places the studies of topological quantum states at one of the frontier topics in condensed matter physics since the last decades~\cite{1,2,3,4}. After a well-understanding on nonmagnetic topological states, breaking the time-reversal symmetry by introducing magnetism in topological materials provides a more fertile playground to give rise to nontrivial topological phases, such as a quantum anomalous Hall effect and axion insulator states~\cite{5,6,7,8}. In magnetic topological materials, the interplay between magnetism and non-trivial band topology can generate new exotic quantum states and will give rise to novel transport phenomena. For instance, magnetic Weyl semimetals Co$_3$Sn$_2$S$_2$~\cite{9}, Mn$_3$Sn~\cite{10} and GdPtBi~\cite{11} have been reported to show a large intrinsic anomalous Hall effect (AHE) originated from the net Berry curvature around the Weyl nodes. On the other hand, a real-space Berry phase arising from a skyrmion phase or a noncollinear spin texture with nonzero scalar spin chirality [$\chi_\textrm{s} = \bm{S}_i \cdot ( \bm{S}_j \times \bm{S}_k) \neq 0$, where $\bm{S}_i$, $\bm{S}_j$, and $\bm{S}_k$ are three nearest spins] can act as a fictitious magnetic field on the conduction electrons, giving rise to topological Hall effect (THE)~\cite{12,13,14,15}. For example, an intrinsic THE arising from a noncollinear spin structure under a magnetic field has been established in the bulk antiferromagnetic (AFM) topological insulator MnBi$_2$Te$_4$~\cite{16}. However, in contrast to the topology in momentum space, electromagnetic responses caused by the noncollinear spin texture in real space need further study.

\begin{figure*}
\includegraphics[width=0.9\textwidth]{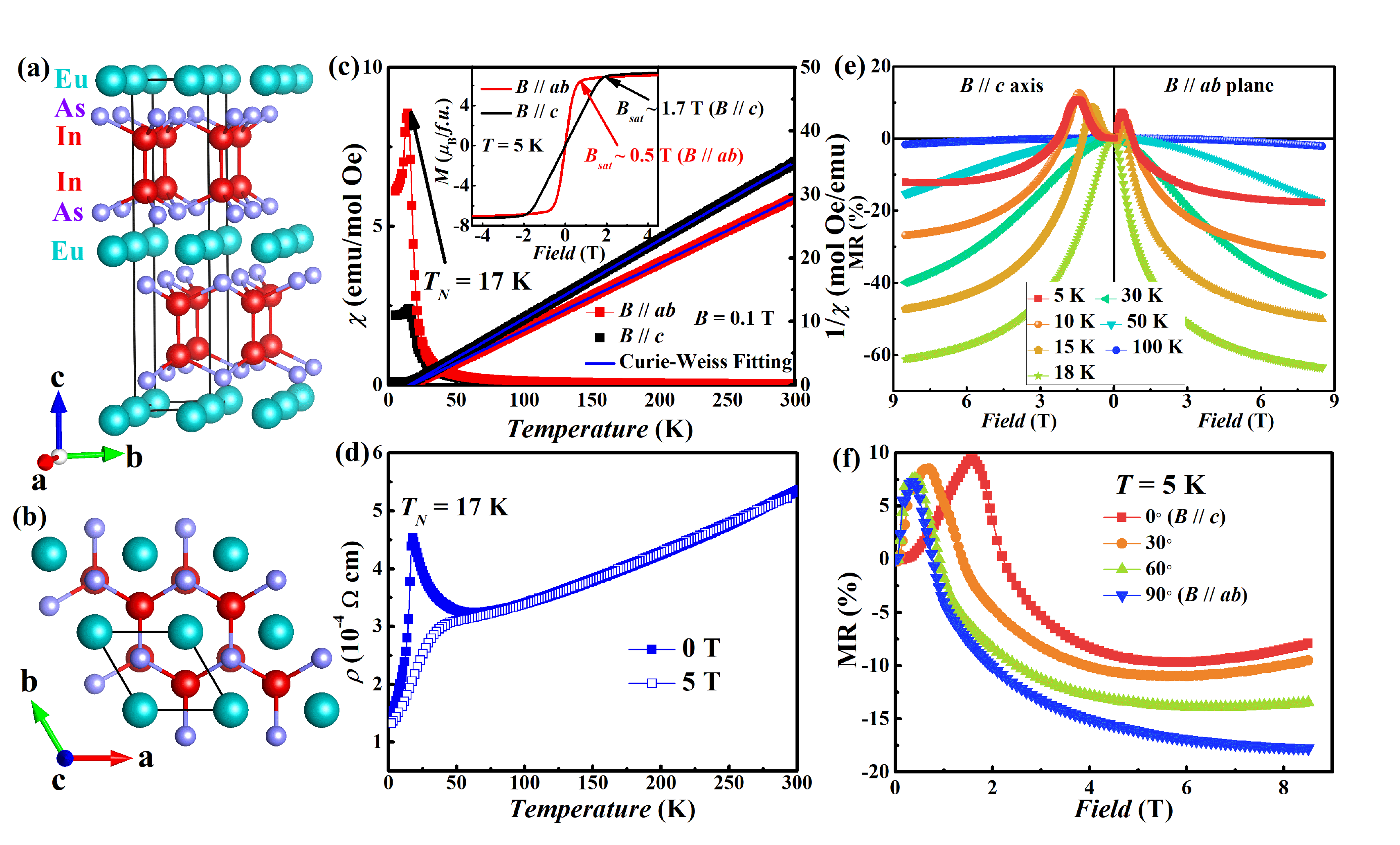}
\caption{The crystal structure of EuIn$_2$As$_2$, viewed along the $a$ axis(a) and the $c$ axis(b). (c) The temperature dependence of the magnetic susceptibility ($\chi$($T$), left) and that of the inverse magnetic susceptibility ($1/\chi(T)$, right) under a magnetic field $B$ = 0.1\,T applied parallel to the $c$ axis (black) and the $ab$ plane (red). The blue lines show a Curie-Weiss fit for the data. The inset shows the field dependence of the magnetization $M(B)$ at 5 K and at $B \parallel c$ (black) and $B \parallel ab$ (red).
(d) The temperature dependence of the resistivity with the current applied along the $ab$ plane at 0\,T and 5\,T ($B \parallel c$).
(e) The field dependence of the magnetoresistance (MR) at various temperatures with $B \parallel c$ (left panel) and $B \parallel ab$ (right panel). (f) The field dependence of MR in different tilted angles from the $c$ axis to the $ab$ plane at 5\,K. }
\label{fig:Graph1}
\end{figure*}

Rare-earth Zintl compound EuIn$_2$As$_2$ is a new promising candidate of the AFM topological materials. EuIn$_2$As$_2$ crystallizes in hexagonal $P$6$_3$$/mmc$ space group with the alternating stacking of Eu and In$_2$As$_2$ layers~\cite{17} as shown in Fig.\,$\ref{fig:Graph1}$(a) and (b), realizing a topological transport of conducting electrons in In$_2$As$_2$ layers affected by the magnetic moments of Eu$^2$$^+$ ions. In fact, a bulk axion insulator phase has been predicted when the Eu moments exhibit the A-type AFM order, in which the ordered Eu moments are ferromagnetically aligned in the $ab$ plane but antiferromagnetically aligned in adjacent layers along the $c$ axis~\cite{17}. Furthermore, different topological states are suggested to be realized depending on the direction of the magnetic moments. A topological crystalline insulator phase with gapless surface states on the (100), (010), (001) surfaces is expected in EuIn$_2$As$_2$ for in-plane oriented Eu magnetic moments. On the other hand, a high-order topological insulator phase with chiral hinge states is predicted with out-of-plane magnetic order~\cite{17}. These different topological states expected in the different orientation of the Eu moments provide a promising way to control the topological state by external magnetic field. Angle-resolved photoemission spectroscopy (ARPES) results show an inversion of the bulk band as entering to the AFM state, suggesting that there is a topological phase transition accomplished with the magnetic transition~\cite{18,19}. Although, recent neutron diffraction experiments show that EuIn$_2$As$_2$ has a low-symmetry helical AFM order, not an A-type AFM order, it has also been predicted that the direction of a modest applied magnetic field of $B \approx 1$--2\,T can tune between gapless and gapped surface states~\cite{20}. Thus, the fruitful topological nature and easily magnetic field tunability of this system will be a promising platform to realize ``clean'' topological quantum effect and many novel transport phenomena.

In this paper, we perform a systematic study of single-crystal samples of EuIn$_2$As$_2$, focusing on the interplay between the topological transport phenomena and the magnetic state. The non-monotonic field dependence of the magnetoresistance (MR) reveals that the transport phenomena is strongly influenced by the spin configuration in EuIn$_2$As$_2$ system. Most importantly, we observe an AHE and a large THE in the AFM state.
From the dependence of the AHE on the longitudinal conductivity and the temperature, we find that the AHE is dominated by the intrinsic mechanism. In addition, from the field and the angle dependence of the THE, we suggest that the THE is attributed to a finite $\chi_\textrm{s}$ caused by the noncoplanar spin texture formed under out-of-plane magnetic field.
%{\color{red}We suggest that the origin of AHE is dominated by intrinsic mechanism. By tuning the magnetic field orientation, we confirm that the THE is attributed to the chiral effect derived from noncoplanar spin texture induced by the $c$-axis component magnetic field.} Our findings may push the advanced experiments to understand the potential physics of this material.
%Our findings may push the advanced experiments to understand the potential physics of this material.

\section{Experimental details}
EuIn$_2$As$_2$ single crystals were grown by the self-flux method. Eu, In and As with a mole ratio 1:12:3 were weighted and ground into alumina crucible in an Ar-filled glove box, which was sealed in a quartz ampoule under high vacuum. The sealed quartz ampoule was heated to 1323\,K and maintained for 12 hours, then cooled down to 873\,K over 100 hours. At this temperature, the quartz ampoule was quickly taken out from the furnace and decanted with a centrifuge to separate EuIn$_2$As$_2$ single crystals. As shown in Fig.\,S1 in the Supplementary Material~\cite{21}, the crystal was grown like a hexagonal piece, with dimensions of 2$\times$2$\times$0.1\,mm$^3$. The X-ray diffraction shows sharp peaks on the pattern of (00$l$) plane, indicating the high-quality single crystallinity of the samples.

Magnetization and electrical transport measurements were carried out by using a Quantum Design Magnetic Property Measurement System (MPMS-XL5), a Physical Properties Measurement System (PPMS-9\,T) for 5\,K\,$< T <400$\,K and $B<9$\,T, and a dilution refrigerator (DR) for 0.1\,K\,$<T<30$\,K and $B <14$\,T. Standard four-probe and five-probe methods were used for the longitudinal resistivity and the Hall measurements with a current in the $ab$ plane, respectively. All the Hall resistivity were obtained by $\rho_{yx}(B)=(\rho_{yx}(+B)-\rho_{yx}(-B))/2$ to remove the influence of misalignment of the contacts. The reproducibility of our results were confirmed by the measurements done for other samples from different batches as shown in Fig. S2 and S3 in the Supplementary Material~\cite{21}.

%\begin{eqnarray}
%\mathcal{\tilde{F}}&=&\alpha_6^*a_1b_2+\alpha_1^*(a_1^2+c_2^2)+\alpha_3^*d_3^2+\alpha_{11}^*(e_1^4+f_2^4)
%\label{eq:energy}
%\end{eqnarray}

\section{Results and discussion}  %This should be cancelled for PRL and APL.

Figure\,$\ref{fig:Graph1}$(c) shows the temperature dependence of the magnetic susceptibility ($\chi$) under a magnetic field ($B$) of 0.1\,T applied along the $c$ axis ($\chi_c$) and the $ab$ plane ($\chi_{ab}$). As shown in Fig.\,$\ref{fig:Graph1}$(c), $\chi_{ab}$ is larger than $\chi_c$ in the whole temperature range we measured, showing the easy axis is in the $ab$ plane.  The AFM transition at N\'eel temperature $T_\textrm{N}=17$\,K is clearly seen by the peak in $\chi$($T$) for both $B \parallel c$ and $B \parallel ab$. The temperature dependence of $\chi^{-1}$ above $T_\textrm{N}$ follows the Curie-Weiss law with a positive Weiss temperature of 16.9\,K and 18.5\,K for $\chi_c$ and $\chi_{ab}$, respectively, showing a dominant ferromagnetic (FM) interaction in EuIn$_2$As$_2$. This dominant FM interaction is suggested to be caused by the magnetic polarons in this material~\cite{22,23}, which reflects the strong exchange interaction between the spins of conduction electrons and the localized Eu moments in EuIn$_2$As$_2$.
%{\color{red}This dominant FM interaction and the AFM transition with the strong anisotropy suggest the formation of the magnetic polarons in this material~\cite{22}, which reflects the strong exchange interaction between the carrier spin and localized Eu spins in EuIn$_2$As$_2$.}
The effective magnetic moment obtained by the Curie-Weiss fit ($8.0 \mu_B$ and 8.7$\mu$$_B$ for $\chi_c$ and $\chi_{ab}$, respectively) agrees well with that expected for the highest spin state of a Eu$^{2+}$(7.94$\mu_B$). The field dependence of the magnetization ($M$) at 5\,K (inset of Fig.\,$\ref{fig:Graph1}$(c)) shows a saturation at $B_\textrm{sat}$ $\sim$ 1.7\,T for $B \parallel c$, which becomes smaller ($\sim 0.5$\,T) for $B \parallel ab$ at 5\,K.

Figure $\ref{fig:Graph1}$(d) shows the temperature dependence of the longitudinal resistivity ($\rho$$_{xx}$) at 0 and 5\,T applied along the $c$ axis. As shown in Fig.\,$\ref{fig:Graph1}$(d), $\rho$$_{xx}$ at 0\,T starts to increase below $\sim$ 60\,K as lowering temperature, which is followed by a sharp peak at $T_\textrm{N}$.
This peak in $\rho$$_{xx}$ shows a strong scattering effect on conduction electrons by magnetic fluctuations at 0\,T, as it completely disappears at 5\,T ($>B_\textrm{sat}$).
%This peak in $\rho$$_{xx}$ completely disappears at 5\,T ($>B_\textrm{sat}$), showing a strong scattering effect on conduction electrons by magnetic fluctuations at 0\,T.
The high onset temperature of the magnetic scattering, which is far above $T_\textrm{N}$, shows the presence of a strong magnetic fluctuation above $T_\textrm{N}$ enhanced by the two-dimensional nature of the material. Figure $\ref{fig:Graph1}$(e) shows the field dependence of the longitudinal resistivity $\rho$$_{xx}$ at different temperatures under the magnetic field applied along the $c$ axis and the $ab$ plane. In the paramagnetic state ($T>T_\textrm{N}$), a negative MR is observed up to the highest field, which reaches a maximum of 60$\%$ at $T_\textrm{N}$.  Below $T_\textrm{N}$, a positive MR appears below $B_\textrm{sat}$, which is followed by a negative MR above $B_\textrm{sat}$. This crossover from the positive MR to the negative one becomes lower as the magnetic field is tilted to the $ab$ plane in accordance with the decrease of $B_\textrm{sat}$ as shown in Fig.\,$\ref{fig:Graph1}$(f), indicating that the positive-to-negative crossover behavior in MR is caused by the magnetic scattering effects enhanced by suppressing the AFM order. Therefore, it confirms that the transport phenomena are strongly influenced by the spin configuration. Effects of magnetic polarons formed by the FM interaction and a crossover from weak antilocalization to weak localization above $B_\textrm{sat}$ have been put forward for the increase of $\rho$$_{xx}$ near $T_\textrm{N}$ and the field dependence of MR below $T_\textrm{N}$~\cite{22,23}.

\begin{figure*}
\includegraphics[width=0.8\textwidth]{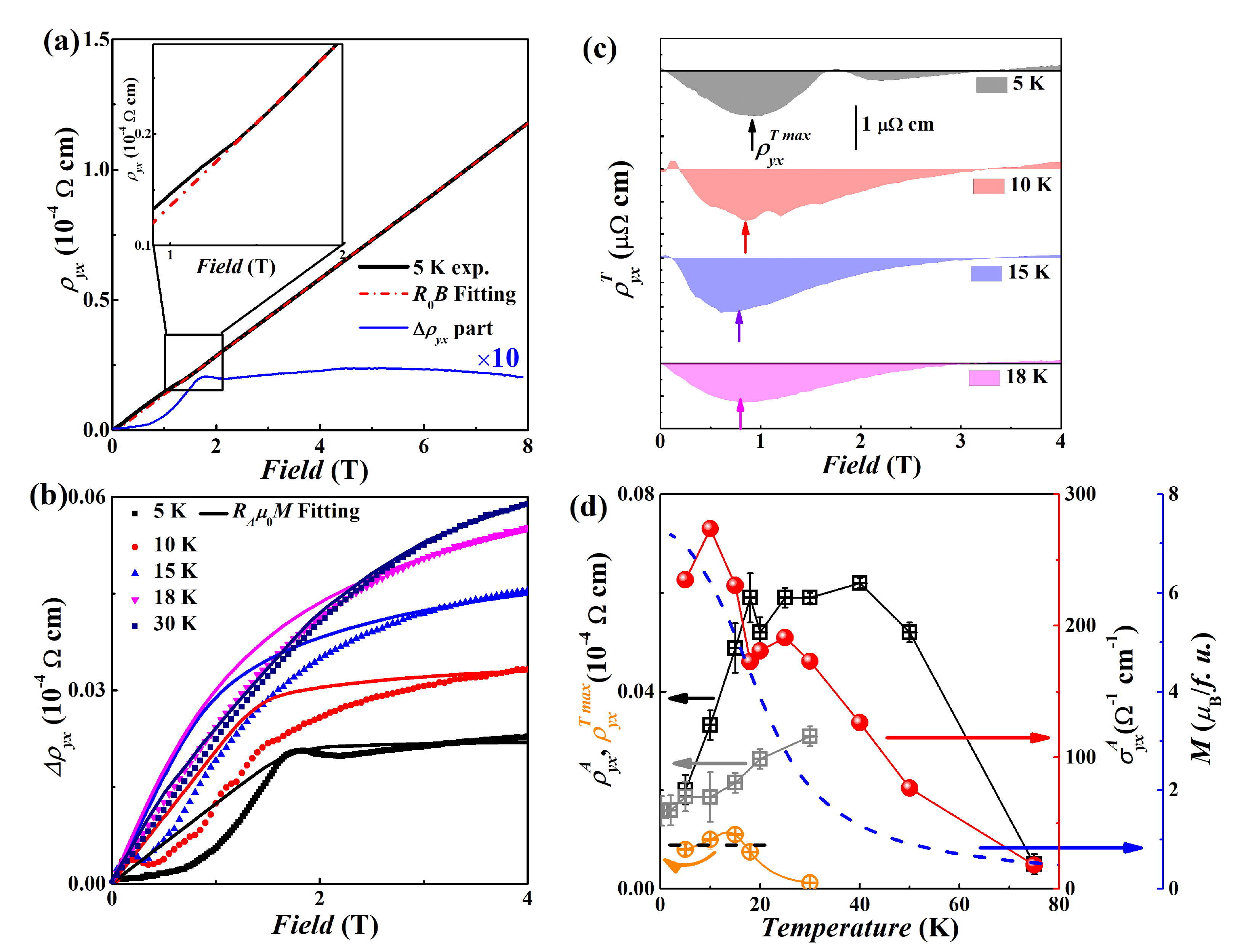}
\caption{Anomalous Hall effect (AHE) and topological Hall effect (THE). (a) The magnetic field dependence of the Hall resistivity ($\rho$$_{yx}$) at 5\,K. The red dashed-dotted line is a fitting curve to estimate the Hall coefficient $R$$_0$, which is used to extract anomalous Hall part $\Delta \rho_{yx} = \rho_{yx} - R_0 B$ (blue line). The inset shows an enlarged view of $\rho$$_{yx}$ for 1--2\,T. (b) The magnetic field dependence of $\Delta$$\rho$$_{yx}$ at different temperatures, obtained after subtracting the normal Hall resistivity. Each solid line shows a fit of the anomalous Hall resistivity $\rho^\textrm{A}_{yx}$ by $\rho^\textrm{A}_{yx} = R_\textrm{A} \mu_0 M$. (c) The magnetic field dependence of the topological Hall resistivity $\rho^\textrm{T}_{yx}$ at various temperatures, the arrows indicate the position of the peak of $\rho^\textrm{T}_{yx}$ ($\rho^\textrm{Tmax}_{yx}$). (d) The temperature dependence of $\rho^\textrm{A}_{yx}$ (black and grey squares for different samples respectively, left axis), $\rho^\textrm{Tmax}_{yx}$ (orange circles, left axis), the anomalous Hall conductivity $\sigma^\textrm{A}_{yx}$ (red circles, 1st right axis), and the magnetization at 2\,T above $B_\textrm{sat}$ (blue dashed line, 2nd right axis).} \label{fig:Graph2}
\end{figure*}

Next, we estimate the AHE in EuIn$_2$As$_2$ system. Figure\,$\ref{fig:Graph2}$(a) shows the field dependence of the Hall resistivity ($\rho$$_{yx}$) at 5\,K (the data of different samples at different temperatures are shown Fig.\,S3 in Supplementary Materials~\cite{21}). As shown in Fig.\,$\ref{fig:Graph2}$(a), $\rho$$_{yx}$ shows a non-linear field dependence near 1.5\,T (see the inset of Fig.\,$\ref{fig:Graph2}$(a)), showing an anomalous contribution in $\rho$$_{yx}$. To determine the field dependence of the anomalous part, $\rho$$_{yx}$ is decomposed to the normal part $\rho^\textrm{N}_{yx} = R_0 B$ and the anomalous part $\Delta \rho_{yx}$ as
\begin{equation*}
\rho_{yx} = \rho^\textrm{N}_{yx} + \Delta\rho_{yx} = R_0 B + \Delta\rho_{yx},
\end{equation*}
where $R$$_0$ is the Hall coefficient. First, we estimate $R$$_0$ by a linear fitting of the field dependence of $\rho$$_{yx}$ above $B_\textrm{sat}$ (the red dashed line) and subtract $\rho^\textrm{N}_{yx}$ from $\rho$$_{yx}$ to obtain the field dependence of $\Delta$$\rho$$_{yx}$ (the bule line in Fig.\,$\ref{fig:Graph2}$(a)).
For the AHE in this material, we estimate that the anomalous Hall resistivity $\rho^\textrm{A}_{yx}$ linearly scales with $M$, which arises from both extrinsic and intrinsic contributions~\cite{16}.
The temperature dependence of $\rho^\textrm{A}_{yx}$ shows a peak at around $T_\textrm{N}$ and decreases as lowering the temperature (the black squares in Fig.\,$\ref{fig:Graph2}$(d)). This decrease of $\rho^\textrm{A}_{yx}$ at lower temperatures continues down to 0.2\,K as shown by the data obtained in a different sample by using DR (the grey squares in Fig.\,$\ref{fig:Graph2}$(d)).

As shown in Fig.\,$\ref{fig:Graph2}$(b), whereas the relation of $\Delta \rho_{yx} \propto M$ holds above $B_\textrm{sat}$, $\Delta$$\rho$$_{yx}$ considerably deviates from $\rho^\textrm{A}_{yx}$ below $B_\textrm{sat}$. This deviation demonstrates an additional component in $\Delta$$\rho$$_{yx}$, named as a topological part $\rho^\textrm{T}_{yx}$. We estimate $\rho^\textrm{T}_{yx}$ by subtracting $R$$_A$$\mu$$_0$$M$ from $\Delta$$\rho$$_{yx}$, where $R$$_A$ is the modified anomalous Hall coefficient and plot the field dependence of $\rho^\textrm{T}_{yx}$ in Fig.\,$\ref{fig:Graph2}$(c). The magnetic field dependence of $\rho^\textrm{T}_{yx}$ at various temperatures below $T_\textrm{N}$ (Fig.\,$\ref{fig:Graph2}$(c)) shows the negative peak at $\sim$ $B_\textrm{sat}$/2. The magnitude of the peak of $\rho^\textrm{T}_{yx}$ ($\rho^\textrm{Tmax}_{yx}$) is almost temperature independent around 1 $\mu\Omega$\,cm, and quickly disappears above $T_\textrm{N}$ (orange circles of Fig.\,$\ref{fig:Graph2}$(d)).

\begin{figure}
\includegraphics[width=0.5\textwidth]{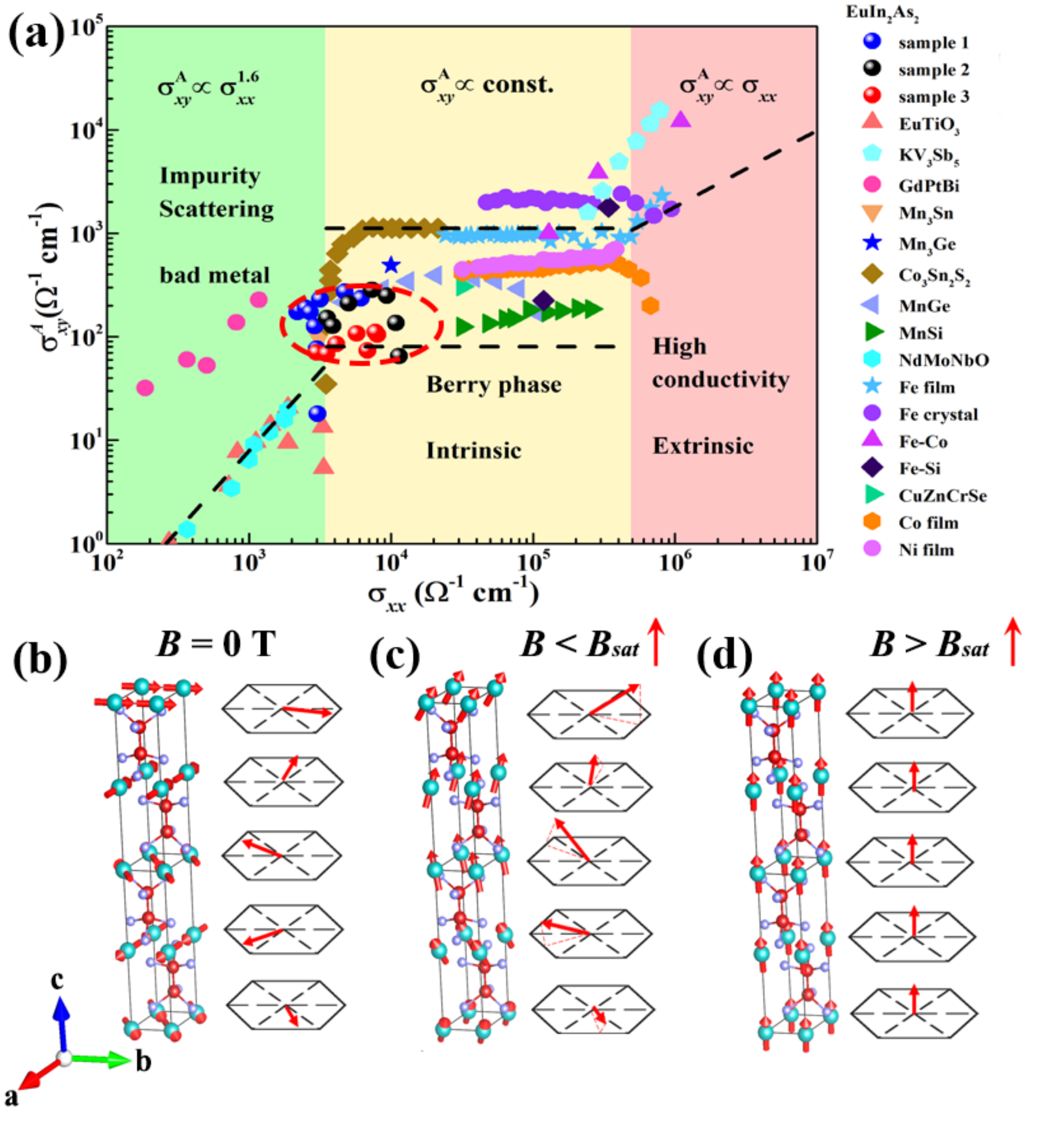}
\caption{(a) Anomalous Hall conductivity $\sigma^\textrm{A}_{yx}$ in the low temperature range  plotted against $\sigma_{xx}$ for EuIn$_2$As$_2$ and other materials in a log-log graph. The ``bad metal regime'', ``intrinsic regime'', and ``extrinsic regime'' of $\sigma_{xx}$ are denoted with different background colors. The reported data for other materials are obtained from refs. ~\cite{9,10,11,31,32,33,34,35,36,37}. (b--d) The spin configurations of Eu with out-of-plane magnetic field. The in-plane helical structure at 0\,T (b) is modified to a noncoplanar structure for $0 < B < B_\textrm{sat}$ (c) and a polarized one for $B_\textrm{sat} < B$ (d).} \label{fig:Graph3}
\end{figure}

We first discuss the origin of $\rho^\textrm{A}_{yx}$ in this compound. As current knowledge, it has been well established that an AHE emerges from the two mechanisms: the intrinsic one caused by a Berry curvature effect~\cite{24,25} and the extrinsic one by either skew or side-jump scattering effects~\cite{26,27}. The former is observed in a moderately dirty metal in which the anomalous Hall conductivity $\sigma^\textrm{A}_{yx} = \rho^\textrm{A}_{yx} / ( {\rho^\textrm{A}_{yx}}^2 + {\rho_{xx}}^2)$ becomes independent on $\sigma_{xx}$, whereas the latter in a clean one in which $\sigma^\textrm{A}_{yx} \propto \sigma_{xx}$~\cite{28}.
In addition, $\sigma^\textrm{A}_{yx}$ of the intrinsic AHE is known to linearly scale to $M$~\cite{24,28}.

We find that the AHE in EuIn$_2$As$_2$ fulfills these features of the intrinsic AHE.
%In the former, the anomalous Hall conductivity $\sigma^\textrm{A}_{yx} = \rho^\textrm{A}_{yx} / ( {\rho^\textrm{A}_{yx}}^2 + {\rho_{xx}}^2)$ becomes independent on $\sigma_{xx}$, whereas the latter by extrinsic mechanism depends on $\sigma_{xx}$.
We plot the anomalous Hall conductivity $\sigma^\textrm{A}_{yx}$ against the longitudinal conductivity $\sigma_{xx}$ with other FM and AFM systems in Fig.\,$\ref{fig:Graph3}$(a). First, we find that the $\sigma_{xx}$ is in the range around 6$\times$10$^3$\,$\Omega^{-1}$cm$^{-1}$ and the $\sigma^\textrm{A}_{yx}$ does not depend on $\sigma_{xx}$. Both features indicate that EuIn$_2$As$_2$ is located at the ``intrinsic regime'', where the intrinsic mechanism is suggested to be dominant.
In addition, as shown in Fig.\,$\ref{fig:Graph2}$(d), we find that the temperature dependence of $\sigma^\textrm{A}_{yx}$ almost follows that of $M$ (i.e. $\rho^\textrm{A}_{yx} \propto \rho_{xx}^2 M$), supporting the dominant intrinsic AHE in $\rho^\textrm{A}_{yx}$. In fact, in the time-reversal-symmetry-broken systems, a band crossing near the Fermi energy $E_\textrm{F}$ with a strong spin-orbit coupling (SOC) can lift the spin degeneracy. The energy gap induced by the SOC produces a large Berry curvature, contributing to the intrinsic AHE. It has been verified that noncollinear antiferromagnets with zero net magnetization can produce a large AHE when their electronic structure exhibits a nonvanishing Berry curvature that acts like a large fictitious magnetic field~\cite{10,29,30}. According to the recent neutron diffraction experiment, EuIn$_2$As$_2$ has a helical magnetic structure~\cite{20}, in which the magnetic moment ferromagnetically aligning in the $ab$ plane rotates antiferromagnetically along the $c$ axis with the magnetic space group of $P$6$_1$2'2'/$C$2'2'21 (Fig.\,$\ref{fig:Graph3}$(b)). These symmetries break both the space and time reversal symmetries, giving rise to local nonzero Berry curvature. In fact, the unitary operators (space symmetry, such as the 6$_1$) and the anti-unitary operators (such as $T$$\cdot$2$_x$) can own a finite Berry curvature in the reciprocal space, ensuring that AHE can emerge in the helical AFM of EuIn$_2$As$_2$.

\begin{figure*}
\includegraphics[width=0.9\textwidth]{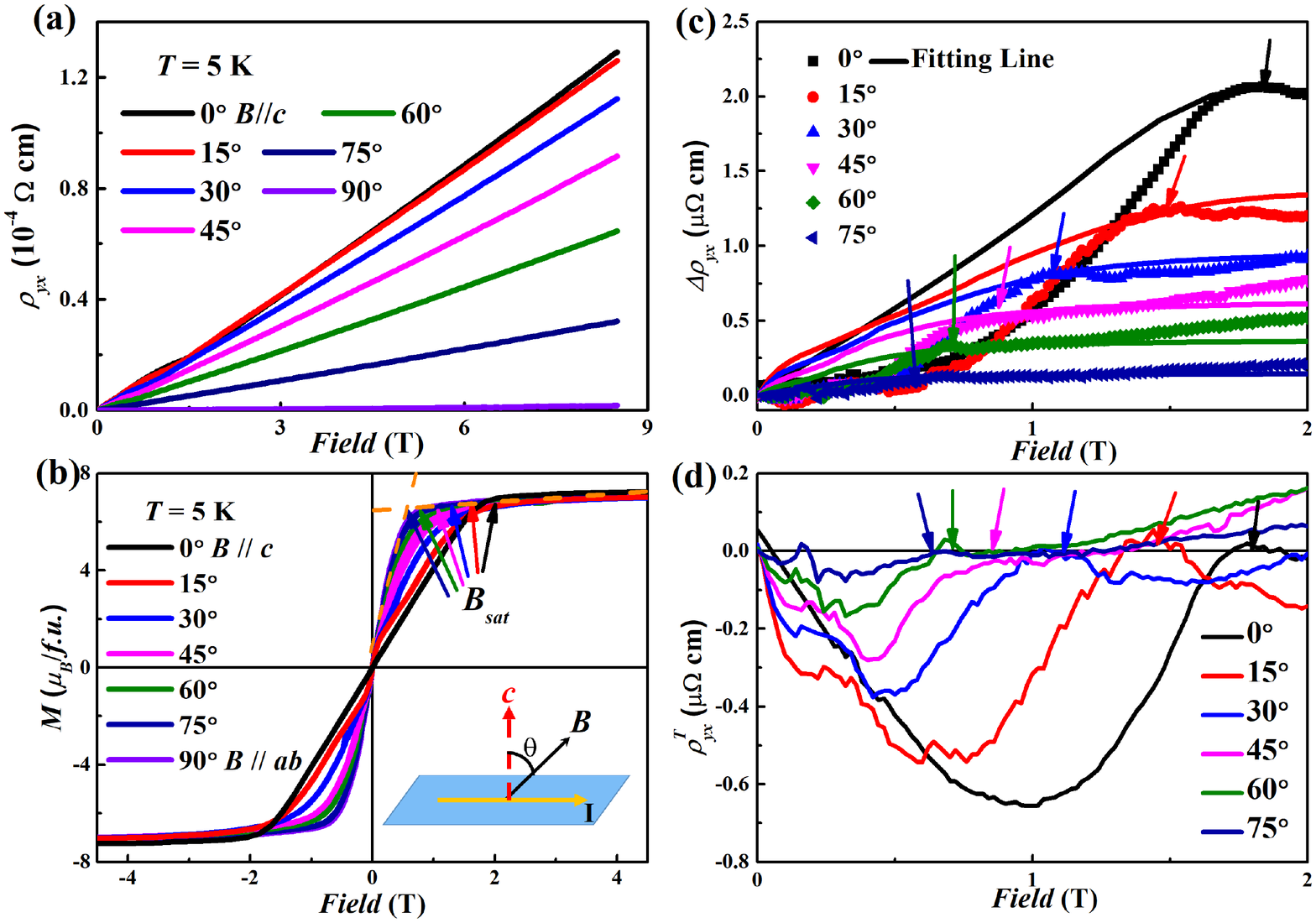}
\caption{(a) The field dependence of the Hall resistivity $\rho_{yx}$ of EuIn$_2$As$_2$ at various tilted angles at 5\,K. (b) The field dependence of the magnetization $M$ at the same angles of (a). (c) The magnetic field dependence of the anomalous Hall resistivity $\Delta$$\rho$$_{yx}$ under various field orientations. Each solid line shows a fitting of $\rho^\textrm{A}_{yx}$ using the relation $\rho^\textrm{A}_{yx} = R_\textrm{A} \mu_0 M$. (d) The magnetic field dependence of the topological Hall resistivity $\rho^\textrm{T}_{yx}$ at various field tilted angles. The arrows in (b), (c) and (d) indicate the saturation field in different tilted angles.} \label{fig:Graph4}
\end{figure*}

Having established the origin of the AHE, we now discuss the origin of $\rho^\textrm{T}_{yx}$. As shown in Fig.\,$\ref{fig:Graph2}$(d), $\rho^\textrm{Tmax}_{yx}$ emerges below $T_\textrm{N}$ and stays a constant at lower temperatures. This temperature dependence is in sharp contrast to $\rho^\textrm{A}_{yx}$ that decreases below 20\,K, suggesting a different origin of $\rho^\textrm{T}_{yx}$. The field dependence of $\rho^\textrm{T}_{yx}$ (Fig.\,$\ref{fig:Graph2}$(c)) shows that $\rho^\textrm{T}_{yx}$ is caused by a deformation of the magnetic structure from either the zero-field state and the fully polarized state.
Since no magnetic transition is observed for $B<B_\textrm{sat}$, the helical magnetic structure at $B=0$ (Fig.\,$\ref{fig:Graph3}$(b)) is continuously deformed to a noncoplanar structure (Fig.\,$\ref{fig:Graph3}$(c)) under a magnetic field applied along the $c$ axis, which is followed by the polarized one above $B_\textrm{sat}$ (Fig.\,$\ref{fig:Graph3}$(d)). Although it remains as a future work to confirm the actual magnetic structure under the magnetic field, this noncoplanar magnetic structure gives rise to a finite $\chi_\textrm{s}$ which eventually vanishes in the fully polarized state above $B_\textrm{sat}$. Therefore, the field dependence of $\rho^\textrm{T}_{yx}$ is well explained by that of $\chi_\textrm{s}$ that develops in the helical magnetic structure under $B \parallel c$~\cite{31,38,39}.
%Under a magnetic field applied along the $c$ axis, the helical magnetic structure is deformed to a noncoplanar one, giving rise to a finite $\chi_\textrm{s}$ (Fig.\,$\ref{fig:Graph3}$(c)) which eventually vanishes in the fully polarized state above $B_\textrm{sat}$ (Fig.\,$\ref{fig:Graph3}$(d)). Therefore, the field dependence of $\rho^\textrm{T}_{yx}$ is well explained by that of $\chi_\textrm{s}$ that develops in the helical magnetic structure under $B \parallel c$~\cite{28,37,38}.
A similar THE is observed in the kagome AFM YMn$_6$Sn$_6$ with the helical magnetic ground state, where the formation of double-fan spin structure under magnetic field contributes to the THE~\cite{40}.
We note that a possibility of a field-induced band structure changing, as observed in EuO~\cite{41}, can be safely excluded as the origin of THE in this material. The band structure calculation of EuIn$_2$As$_2$ predicts that the energy shift of the unoccupied band closes the inverted band gap near the $\Gamma$ point when the magnetic order is changed from helical to FM, suggesting a possibility of the band-structure control by applying a magnetic field~\cite{20}. However, in contrast to these band calculations predicting a semiconducting state for EuIn$_2$As$_2$, this compound is a naturally hole-doped metal as shown by our Hall measurements. In fact, the previous ARPES measurements~\cite{18} show that the actual Fermi energy is far below the energy gap ($\sim -0.3$~eV), indicating that the slight energy shift of the unoccupied band does not affect the electric transport of this material.

To confirm this topological transport caused by $\chi_\textrm{s}$ of the noncoplanar magnetic structure in EuIn$_2$As$_2$, we investigate the angle dependence of $\rho^\textrm{T}_{yx}$ by tilting the magnetic field from the $c$ axis to the $ab$ plane. Figure\,$\ref{fig:Graph4}$(a) shows the field dependence of $\rho$$_{yx}$ at different angles at 5\,K. Simultaneously, we also take the magnetization data at the same tilted angle of the magnetic field, displayed in Fig.\,$\ref{fig:Graph4}$(b). First, we extract the field dependence of $\Delta$$\rho$$_{yx}$(Fig.\,$\ref{fig:Graph4}$(c)) by subtracting the normal part. As shown in Fig.\,$\ref{fig:Graph4}$(c), $\Delta$$\rho$$_{yx}$ is gradually suppressed by tilting the magnetic field to the $ab$ plane. Then we focus on the THE below the saturation field in different tilted angle, we obtain $\rho^\textrm{T}_{yx}$ in the tilted magnetic field by subtracting the $M$-linear component from $\Delta$$\rho$$_{yx}$. As shown in Fig.\,$\ref{fig:Graph4}$(d), $\rho^\textrm{T}_{yx}$ decreases as the magnetic field is tilted to the $ab$ plane.
This angle dependence indicates that $\chi_\textrm{s}$ caused by the $c$-axis component of the applied field is directly linked to the THE.

Similar AHEs are also observed in the related Eu-based compounds EuCd$_2$Sb$_2$~\cite{42} and EuCd$_2$As$_2$~\cite{43}. In these compounds, the Weyl points near the Fermi energy or a dynamical Weyl semimetal state are suggested to contribute the AHEs. It should be noted that, although a similar THE might be observed in these compounds below the saturation field, the origin of the THE is not discussed in these previous works~\cite{42,43}. Given the different collinear AFM state suggested in EuCd$_2$Sb$_2$ and EuCd$_2$As$_2$ and the different crystal structure of these compounds, the origin of the THE in EuCd$_2$Sb$_2$ and EuCd$_2$As$_2$ may be different from that in EuIn$_2$As$_2$.

\section{Summary}  %This should be cancelled for PRL and APL.

In conclusion, our detailed transport measurements on bulk EuIn$_2$As$_2$ single crystals reveal that the transport properties in this compound are strongly influenced by the spin texture formed below $T_\textrm{N}$. Most importantly, we find an AHE and a THE emerge in the AFM state. We suggest that the AHE is originated from a nonvanishing net Berry curvature in the momentum space due to the helical spin structure and that the THE is attributed to the scalar spin chirality of the noncoplanar spin structure caused by the external field. These observations and analysis in EuIn$_2$As$_2$ provide a fertile ground to understand the influence of the interplay between the topology of electronic bands and field-induced novel magnetic structure on unconventional magnetoelectric response. Our findings may push the advanced experiments to understand the nature of this material and other magnetic topological materials.

\section{Acknowledgements}
This work was supported by the National Nature Science Foundation of China under Contract Nos. 11674326, 11874357, the Joint Funds of the National Natural Science Foundation of China, the Chinese Academy of Sciences' Large-Scale Scientific Facility under Contract Nos. U1832141, U1932217, and U2032215, the Key Research Program of Frontier Sciences, CAS (No. QYZDB-SSW-SLH015), the uses with Excellence and Scientific Research Grant of Hefei Science Center of CAS (No.2018HSC-UE011). The work in Japan was supported by KAKENHI (Grants-in-Aid for Scientific Research) Grants No. JP19H01848 and No. JP19K21842.

%\bibliography{Multiferroic_20070522}% Produces the bibliography via BibTeX.

\end{document}